\documentclass{aa}  

\usepackage[T1]{fontenc}
\usepackage{amsmath}	
\usepackage{soul}
\usepackage{xspace}
\usepackage{subcaption}
\usepackage{gensymb}
\usepackage{bm}	
\usepackage{graphicx}
\usepackage{newtxtext}
\usepackage[smallerops,cmbraces]{newtxmath}
\usepackage{xcolor}
\usepackage{newtxtext,newtxmath}
\usepackage{rotating}
\usepackage{adjustbox}
\usepackage{multicol}

\usepackage{subcaption}

\usepackage[T1]{fontenc}
\definecolor{xlinkcolor}{cmyk}{1,1,0,0}

\usepackage[
  bookmarks=true,         
  pdfnewwindow=true,      
  colorlinks=true,        
  linkcolor=xlinkcolor,   
  citecolor=xlinkcolor,   
  filecolor=xlinkcolor,   
  urlcolor=xlinkcolor,    
  final=true,
]{hyperref}

\usepackage[normalem]{ulem} 

\makeatletter
\renewcommand*\aa@pageof{, page \thepage{} of \pageref*{LastPage}}
\usepackage{natbib}
\bibpunct{(}{)}{;}{a}{}{,}%
\def\bibfont{\aa@bibliographyfont}%
\makeatother

\begin{document}

\title{Built to Rest: The Evolving Star-Forming Main Sequence Requires Episodic Quiescence or Late Assembly}
\titlerunning{Galaxies evolving on the SFMS}

\author{
    Lucas C. Kimmig\inst{\ref{inst:usm},\ref{inst:nottingham}\thanks{E-mail: lkimmig@usm.lmu.de}}
    \and
    Jesse van de Sande\inst{\ref{inst:unsw}}
    \and
    Rhea-Silvia Remus\inst{\ref{inst:usm},\ref{inst:swin}}
    \and
    Klaus Dolag\inst{\ref{inst:usm},\ref{inst:mpa}}
    \and
    Rebecca Davies\inst{\ref{inst:swin}}
    \and\\
    Deanne Fisher\inst{\ref{inst:swin}}
}
\authorrunning{L. C. Kimmig et al.}

\institute{
    Universitäts-Sternwarte, Fakultät für Physik, Ludwig-Maximilians-Universität München, Scheinerstr. 1, 81679 München, Germany\label{inst:usm}
    \and 
    School of Physics and Astronomy, University of Nottingham, University Park, Nottingham NG7 2RD, UK\label{inst:nottingham}
    \and
    School of Physics, University of New South Wales, Sydney,  NSW 2052, Australia\label{inst:unsw}
    \and 
    Centre for Astrophysics and Supercomputing, Swinburne University of Technology, Hawthorn VIC 3122, Australia\label{inst:swin}
    \and
    Max-Planck-Institut für Astrophysik, Karl-Scharzschild-Str. 0, 85748 Garching, Germany\label{inst:mpa}
}

\date{Received XX Month, 20XX / Accepted XX Month, 20XX}

\abstract
{The star-forming main sequence of galaxies has now been observed out to redshifts of $z\sim6$ and beyond. However, it remains unclear how long typical galaxies remain on or near it as they evolve, and how frequently they return after departing from it. To determine the expected star formation histories, we construct an analytical model to evolve galaxy properties along the star-forming main sequence over time. Our modeled star formation histories and mean ages agree remarkably well with those reconstructed from observational data. Older and more peaked star formation histories arise naturally for more massive galaxies. Simultaneously, we demonstrate that low-mass ($M_*\geq10^{8}\mathrm{M}_\odot$), early-forming ($z>3$) progenitors that remain on the star-forming main sequence must evolve into very massive ($M_*\approx10^{11}\mathrm{M}_\odot$) galaxies today. Consequently, the progenitors of intermediate mass galaxies ($M_*=10^{10}\mathrm{M}_\odot$) must have either formed late ($z<2$) or underwent significant phases ($T>1\,$Gyr) with suppressed star formation rates ($0.3\,$dex below the star-forming main sequence). We provide tracks to connect galaxies from $z=6$ to $z=0$ by their star-forming behavior above or below the main sequence. By applying number density arguments to construct evolutionary histories for Milky Way-mass galaxies, we find that they must undergo a significant phase of suppressed star formation, nearing quiescence, or otherwise become too massive. This is particularly true for the Milky Way itself, where we show that the observed presence of a large amount of old stars directly implies a departure from the star-forming main sequence over the majority of its history.}


\maketitle

\section{Introduction}\label{sec:intro}
Galaxies of all masses are known to grow either by forming stars themselves, i.e. in-situ, or through mergers with another galaxy \citep[e.g. ex-situ,][]{oser:2010,remus:2022}. Those galaxies that are actively forming stars do so in a manner that establishes a relation between their star formation rates and their current stellar masses, resulting in the so-called star-forming main sequence (SFMS) of galaxies. This SFMS is observed at redshifts as high as $z=6$ and down to $z=0$, with the relation evolving with decreasing redshift toward lower star formation rates at a given stellar mass \citep[e.g.,][]{speagle:2014,pearson:2018,leslie:2020,popesso:2023}. However, while this behavior of the relation is consistently observed, the normalization and slopes of the relation deviate between observational campaigns \citep[e.g.,][]{popesso:2023,fortune:2025}, and precisely which physical processes dominate in establishing this correlation are still debated.

Disk galaxies are known to populate the star-forming main sequence over a large range of stellar masses at $z=0$, from galaxies as small as M33 and NGC~300 with about $M_*\approx10^{9}\mathrm{M}_\odot$ \citep[e.g.,][for galaxies in the local volume]{argudo:2025} to the most massive disk galaxies in the present-day Universe, the super spirals with masses that can even reach $M_*\approx1\times10^{12}\mathrm{M}_\odot$ \citep{diteodoro:2021,ogle:2019}. The typical range in star formation rate (SFR) used in the literature to define galaxies as lying on the star forming main sequence is about~$\pm0.3\,$dex \citep{noeske:2007,whitacker:2012,speagle:2014,pearson:2018}. Deviations from the SFMS towards higher star formation rates are usually correlated to (gas-rich) merger events \citep[e.g.,][]{genzel:2010,gomez_guijarro:2022}, while deviations towards lower star formation rates are thought to be indicative of ongoing quenching either through mergers \citep[e.g.,][]{ellison:2022} or other mechanisms such as a lack of gas replenishment due to halo mass quenching \citep[e.g.,][]{birnboim:2003,dekel:2008} or feedback from stars and black holes \citep[e.g.][]{kimmig:2025}.

The questions remains whether the SFMS is primarily (1) a tight ($\pm0.3\,$dex) evolutionary pathway or (2) an ensemble relationship formed by galaxies undergoing cycles of quiescence and rejuvenation. In the former case, an evolution along the main sequence for a given galaxy would equate to a continuous accretion of gas from the cosmic web that is balanced by continuous star formation in and gas ejection from the galaxy, a model commonly referred to as the ``bathtub model'' \citep[e.g.,][]{bouche:2010,lilly:2013,burkert:2017}. In this simple picture, the galaxy grows continuously through in-situ star formation in an equilibrium state. This equilibrium may be maintained by galactic outflows that prevent overly high star formation efficiencies \citep{reichardt-chu:2022}, until star formation is ultimately terminated by, for example, a merger event or the build-up of a hot medium \citep[e.g.,][]{white:1991,dekel:2006,tacchella:2016}. Consequently, the decline of the SFMS with decreasing redshift \citep[e.g.,][]{popesso:2023} could then be naturally explained by the drop in gas accretion rate with decreasing redshift \citep[e.g.][]{seidel:2025}, resulting in the lower observed molecular gas fractions of galaxies on the SFMS at later times \citep{tacconi:2020}. In addition, later forming halos possess higher spin \citep{teklu:2015}, such that the gas flows are also of higher angular momentum and would thus be accreted farther out at lower densities, where star formation is consequently less efficient \citep[e.g.,][]{noeske:2007,teklu:2023}. 

If the SFMS is instead primarily an {\it ensemble relationship}, then galaxies are expected to move on and off the relation through time (defined as going beyond $\pm0.3$dex) in a cycle of quiescence and rejuvenation, as is recently being found in simulations \citep{fortune:2025}. In this case, feedback from within the galaxy can temporarily (on the order of~$1\,$Gyr or longer) quench the star formation before accretion from the cosmic web or merging can then bring the galaxy back onto the main sequence \citep{remus:2025}. In fact, \citet{fortune:2025} find that such cycles are rather common events for more than half of the galaxies at $z=0$ with stellar masses above $10^{10}\mathrm{M}_\odot$, with some galaxies running through 6~such cycles in their evolution.
Although challenging to measure, observations are also beginning to reveal indications of such on-off episodes of star formation. For example, \citet{munozlopez:2025} find multiple star formation cycles for~$\sim85\%$ of a set of 393~galaxies from CANDELS/GOOD-S and CANDELS/COSMOS. Similarly, \citet{looser:2024} and \citet{trussler:2025} find indications for such on-off cycles in galaxies at high redshifts before cosmic noon, while \citet{simmonds:2025} find increasing scatter in the SFMS at those redshifts, which may indicate less of a stable equilibrium. Previously, \citet{chauke:2019} had already reported rejuvenation indications in the galaxies from the LEGA-C survey at $z\approx0.8$.

Determining whether galaxies of a given mass follow a preferred star formation history, and whether the SFMS is populated by systems evolving steadily along it, undergoing cycles of quiescence and rejuvenation, or both, requires accurate measurements of the stellar age distribution, which are notoriously difficult.
Determining whether galaxies of a given mass follow a preferred star formation history, and whether the SFMS is populated by systems evolving steadily along it, undergoing cycles of quiescence and rejuvenation, or both, requires accurate measurements of the stellar age distribution, which are notoriously difficult to measure. Instead, when working on the average stellar age of galaxies, \citet{mattolini:2025} report a positive correlation with the stellar mass of SDSS galaxies, especially for $M_*>10^{10}\mathrm{M}_\odot$. A similar trend was already reported for early-type galaxies, for example by \citet{thomas:2010} using Lick indices and by \citet{mcdermid:2015} using pPXF fitting \citep{cappellari:2017}, with both finding average ages older than $10\,$Gyrs for early-type galaxies with more than $M_*>10^{11}\mathrm{M}_\odot$. Beyond those works, other spectral energy distribution (SED) fitting codes, such as Bagpipes \citep{carnall:2018} and Prospector \citep{johnson:2021}, are commonly used to recover the SFH of galaxies, albeit with their own uncertainties.

Resolved stellar population ages can, as of yet, only be obtained for galaxies close to our own Milky Way galaxy, and best for the Milky Way itself. Based on APOGEE data, \citet{haywood:2016} found that our own Milky Way has formed a significant amount of its stellar mass already before $z=2$, after which it fell to a low star formation rate, including a noticeable dip in star formation between $z=1-2$ indicative of a period of quiescence. This drop in star formation rate coincides with the time of the last major merger event of our Milky Way galaxy, Gaia Enceladus, as reported by \citet{helmi:2018} and \citet{belokurov:2018} to be around $z\approx2$. Additional evidence for such an evolutionary scenario was reported based on O-stars by \citet{xiang:2022} as well. However, with only a small sample of galaxies close enough for detailed measurements of individual stars, and thus more accurate star formation histories, it is unclear if such an on-off growth scenario is common for galaxies of Milky Way mass, or if this is a special case. 

Under the assumption that Milky Way-mass galaxies assembled coherently, such that the number density of their progenitors through cosmic time should be conserved, \citet{vandokkum:2013} and \citet{papovich:2015} determined the resulting stellar mass growth history. They predict that Milky Way-mass galaxies should have had masses of $M_*>10^{9}\mathrm{M}_\odot$ already at $z\approx3$, somewhat below the results by \citet{haywood:2016}, with even larger masses at high redshift for Andromeda galaxy mass counterparts. As of yet, it remains unclear if these massive progenitors, such as the barred galaxy CEERS-2112 at $z\approx3$ with a stellar mass larger than $M_*>10^{9}\mathrm{M}_\odot$ observed by \citet{costantin:2023}, could have evolved along the SFMS to become galaxies like the Milky Way today, or would have necessarily grown more massive. 

As the number of observations of massive disks at high redshifts of $z>2$ continues to increase, the question of their possible present-day descendants is becoming increasingly important. Beyond CEERS-2112, \citet{rowland:2024} for example report REBELS-25, a dynamically cold disk at $z=7.3$ with $M_*\approx8\times10^9\mathrm{M}_\odot$, while \citet{wang:2025} and \citet{umehata:2025} reported two different massive disks at $z\approx3$ with $M_*>10^{11}\mathrm{M}_\odot$. Even more extreme, \citet{xiao:2025} find a star forming disk galaxy at $z=5.2$ from the JWST-Panoramic survey with $M_*=2\times10^{11}\mathrm{M}_\odot$. These galaxies are massive enough at high redshifts to be potential progenitors to the super spiral galaxies ($M_*=2\times10^{11}\mathrm{M}_\odot$) reported today \citep{ogle:2019}.

In this study, we consider the formation pathways, average stellar age distributions, and final stellar masses of galaxies that evolve entirely on the star-forming main sequence, based on a model we developed. We introduce the model in Sec.~\ref{sec:model}, and present the resulting properties of main sequence galaxies in Sec.~\ref{subsec:mainseq}. In Sec.~\ref{subsec:connect}, we discuss how to connect galaxies through cosmic time, including evolutionary tracks for present-day stellar masses expected from an observed stellar mass at $z=6,5,4,3,2,1$ in Fig.~\ref{fig:grid_chab} and the appendix Fig.~\ref{figA:grid_chab}. In Sec.~\ref{subsec:mw}, we apply our model to the Milky Way and discuss the resulting implications. Finally, we conclude the paper with a discussion in Sec.~\ref{sec:conc}. All our calculations assume a 0.3/0.7 $\Lambda$CDM concordance cosmology, with $\Omega_m=0.3$, $\Omega_\Lambda=0.7$ and $h_0=0.7$.

\section{The Model}\label{sec:model}

To track the evolution of stellar mass and star formation rate of model galaxies, we need to consider four key criteria.

\begin{enumerate}
    \item The parametrization of the star-forming main sequence (SFMS).
    \item The slope and normalization of the initial mass function (IMF).
    \item The timestep size of the iteration $\delta t$.
    \item The initial progenitor galaxy properties. 
\end{enumerate}

\noindent For our SFMS we select the parametrization given by \citet{popesso:2023} because of its coverage in redshift ($0<z<6$), stellar mass ($M_*=10^{8.5}$ to $10^{11.5}\mathrm{M}_\odot$) and SFR ($0.01$ to $500\mathrm{M}_\odot\mathrm{yr}^{-1}$). In addition, they fit the functional form of their SFR dependent on both the stellar mass and, crucially, the cosmic time directly. This is necessary to smoothly integrate our model galaxies along the SFMS as opposed to redshift bins with discontinuous jumps in SFR. Specifically then, we chose the turnover mass parametrization by \citet{popesso:2023}, where 
\begin{equation}\label{eq:sfms}
    \log\mathrm{SFR}_\mathrm{SFMS}(\mathrm{M_*},t) = (a_0 + a_1t) - \log\left(1 + \left(\frac{M_*}{10^{a_2+a_3t}}\right)^{-1}\right) 
\end{equation}
with $a_0=2.71$, $a_1=-0.186$, $a_2=10.86$ and $a_3=-0.0729$ (their Eq.~14). The term $10^{a_2+a_3t}$ works as a turnover mass $\mathrm{M}_\mathrm{T}(t)$, above which the SFMS flattens, while $10^{a_0 + a_1t}$ is the maximum SFR$_\mathrm{max}(t)$ reached for $M_*\to\infty$. This can be seen directly when re-writing Eq.~\ref{eq:sfms} as:
\begin{equation}\label{eq:sfms_turn}
    \mathrm{SFR}_\mathrm{SFMS}(\mathrm{M_*},t) = \frac{\mathrm{SFR}_\mathrm{max}(t)}{1 + M_\mathrm{T}(t)/M_*}.
\end{equation}

For our IMF we use a \citet{chabrier:2003} mass function, which dictates the fraction and timescale of the formed stellar mass that is lost to winds and supernovae. In particular, we use the parametrization of the fraction of stellar mass loss (with remnant mass in BHs, neutron stars not counted as mass loss) as a function of time given by \citet{leitner:2011}, 
\begin{equation}
    f_\mathrm{ml}(t)=C_0\cdot\ln(t/\lambda+1),
\end{equation}
with $C_0=0.046$ and $\lambda=2.76\times10^{5}\, \mathrm{yr}$. As only the high-mass slope really alters the loss rate \citep{leitner:2011}, our findings hold true for IMFs with similar slopes in that regime, such as the \citet{kroupa:2001} IMF. Conversely, a \citet{salpeter:1955} IMF would result in a significantly lower stellar mass loss rate, which would consequently accelerate the evolution of our model galaxies toward higher stellar masses. Our model stellar masses can thus be understood as lower limits for IMFs with steeper high-mass slopes.

The timestep size defines how often the stellar mass and SFR of our model galaxies are updated, where we choose $\delta t = 100\,\mathrm{Myr}$. We have tested values from $\delta t = 5\,\mathrm{Myr}$ to $1\,\mathrm{Gyr}$, and find that the results converge within $5\%$ for $\delta t < 200\,\mathrm{Myr}$ when the galaxy remains at a fixed logarithmic separation to the star-forming main sequence through cosmic time (i.e., $\log{\mathrm{SFR}(M_*,t)} = \log{\mathrm{SFR}_\mathrm{SFMS}(\mathrm{M_*},t)} + \Delta_\mathrm{SFMS}$, with $\Delta_\mathrm{SFMS}$ fixed). Differences between the shorter timesteps become more pronounced (though remain at $< 15\%$) when allowing for a lognormal random scatter around the SFMS (i.e., $\Delta_\mathrm{SFMS}=\Delta_\mathrm{SFMS}(t)$), which we discuss in more detail in appendix~\ref{app:timesteps}. Consequently, we chose $\delta t = 100\,\mathrm{Myr}$, which is also the age of the stellar populations that contribute $90\%$ of the emission to most SFR tracers except H$\alpha$ \citep{kennicutt:2012}. 

Finally, the progenitor galaxy of stellar mass $M_1$ at cosmic time $t_1$ (or equivalently redshift $z_1$) should itself be attributed an age, so that it also experiences mass loss with time, as opposed to having a ``static'' core of mass $M_1$. This choice of age is a priori arbitrary, with the only requirement being that the original mass $M_0$ seeded at time $t_0\leq t_1$ results in the ``observed'' mass $M_1$ at $t_1$, i.e., $M_\mathrm{1} = M_0 \cdot\left[1-f_\mathrm{ml}(t_1-t_0)\right]$. For example, if $t_0=t_1$ then $M_0=M_1$ as $f_\mathrm{ml}(0)=0$. We chose $t_0=917\,\mathrm{Myr}$ for all progenitors, equivalent to $z_0=6$ in our concordance cosmology. To mitigate the impact of the unavoidable discontinuity in the star formation history at whichever chosen $z_0$, we set our progenitor stellar masses low at $M_*(t=t_1) = M_1=1\times10^{8}\mathrm{M}_\odot$. This is low enough to be just $1\%$~of the final mass for galaxies with $M_*>1\times10^{10}\mathrm{M}_\odot$, while remaining near the $M_*\approx3\times10^{8}\mathrm{M}_\odot$ lower bound of the SFMS fit by \citet{popesso:2023}. 

Our final algorithm then looks as follows:
\begin{enumerate}
    \item Input the progenitor stellar mass of $M_1$ at time $t_1$ and determine its age relative to $z=6$.
    \item Iteration from timestep $n=1$ to $n=N$:
    \begin{itemize}
        \item{Apply mass loss to stellar mass bins $m_i$ from $i=1$ to $i=n$ according to $f_\mathrm{ml}(t_n - t_i)$}.
        \item{Determine total current mass $M_n=\sum_{i=1}^n m_i$ and resulting $\mathrm{SFR}_n(M_n,t)$}.
        \item{Add next stellar mass bin $m_{n+1} = \mathrm{SFR}_n(M_n,t_n)\times \delta t$ with birth time $t=t_{n+1}$}.
    \end{itemize}
    \item Output list of total masses $M_n$, star formation rates $\mathrm{SFR}_n$, and corresponding steps in cosmic time $t_n$.
\end{enumerate}
\noindent
Note that at timestep $n$ the stellar mass $m_n$ that was just formed is taken to have a mean stellar age of half of the timestep size $\delta t/2$, which effectively means that the stellar mass is added at the middle of the timestep. This choice means that a portion of the stellar mass formed within the timestep is immediately lost to supernovae, as for our timestep size $\delta t/2=50\,\mathrm{Myr}$ we have $f_\mathrm{ml}=0.24$ with a Chabrier IMF. This is very close to the mean mass loss that results from subdividing each $100\,\mathrm{Myr}$ timestep into $1\,\mathrm{Myr}$ blocks, which gives $f_\mathrm{ml}=0.23$, meaning that our choice well approximates a steady formation of stars over $100\,\mathrm{Myr}$. The minor difference becomes smaller with increasing age of the stars.

\section{Results}\label{sec:results}

\subsection{Properties of main sequence galaxies}\label{subsec:mainseq}
\begin{figure}
\centering
\includegraphics[width=0.9\columnwidth]{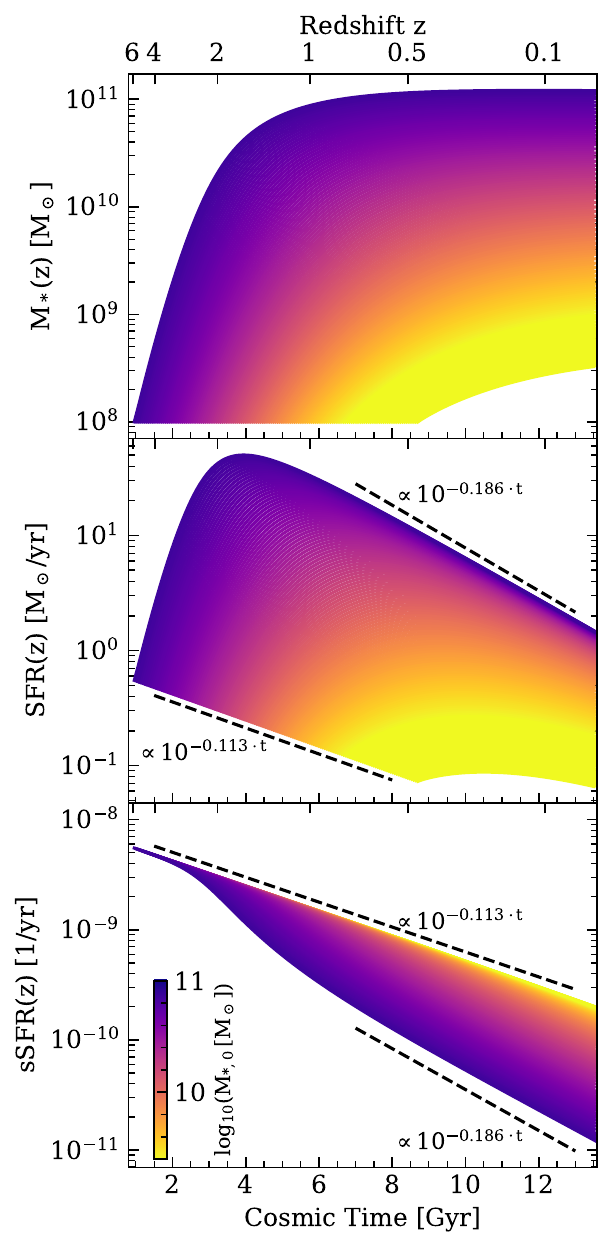}
\caption{Stellar mass (M$_*$, \textbf{top}), star formation rate (SFR, \textbf{middle}), and specific star formation rate (sSFR, \textbf{bottom}) against cosmic time for galaxies evolving along the \citet{popesso:2023} main sequence. Progenitors are seeded between $z=6$ and $z=0.5$ at $M_*=10^{8}\mathrm{M}_\odot$, while lines are colored by their final $z=0$ stellar mass, $\mathrm{M}_{*,z=0}$. We find that with increasing $\mathrm{M}_{*,z=0}$, the peak SFR shifts to earlier times and higher values. At any redshift, the sSFR of more massive galaxies is lower, while for a given stellar mass, the sSFR decreases with time.}\label{fig:intro}
\end{figure}

First, we consider how a broad range of galaxies would evolve if they remain fixed on the SFMS. To this end, we seed progenitors with $M_\mathrm{*,z}=1\times10^{8}\mathrm{M}_\odot$ from $z=0.5$ to $z=6$ in $150$~equal steps in cosmic time (a spacing of $50\,\mathrm{Myr}$) and plot the evolution of their stellar mass (top, $M_*$), star formation rate (middle, SFR), and specific star formation rate (bottom panel, $\mathrm{SFR}/M_*$) as a function of cosmic time in Fig.~\ref{fig:intro}. We color each line by the final stellar mass, as indicated in the colorbar. As expected, seeding the progenitor at earlier times results in larger $z=0$ stellar masses, $\mathrm{M}_{*,z=0}$. We determine this dependence to follow a linear relation with $\log{\mathrm{M}_{*,z=0}} = 11.43 - 0.36\cdot t$, where $t$ is the cosmic time in Gyr of the progenitor. This relationship is remarkably tight, with a median absolute deviation in $\log{\mathrm{M}_{*,0}}$ of just $0.023\,$dex ($\approx5.4\%$). 

As can be seen from the lower panel, at a given redshift, the \citet{popesso:2023} SFMS finds more massive galaxies to have lower sSFR, but notably also that galaxies of equal stellar mass decrease their sSFR with increasing cosmic time. This is apparent from the upper bound of the curves, which represents the sSFR of the progenitor galaxies that all start with $\mathrm{M}_\mathrm{*,z}=1\times10^{8}\mathrm{M}_\odot$ and which follows a slope of sSFR$\propto10^{-0.113\cdot t}$, indicated by the upper dashed line. 

This slope directly follows from the turnover mass parametrization of the \citet{popesso:2023} SFMS. For low stellar masses $M_*\ll \mathrm{M_\mathrm{T}(t)}$ (i.e., $1\ll \mathrm{M_\mathrm{T})(t)}/M_*$, Eq.~\ref{eq:sfms_turn} becomes
\begin{align*}
    \mathrm{SFR}(\mathrm{low\, M}_*,t)&\approx M_*\cdot\mathrm{SFR}_\mathrm{max}(t)/\mathrm{M}_\mathrm{T}(t)\\
    &=M_*\cdot10^{a_0+a_1t - (a_2+a_3t)} 
    =\frac{M_*}{10^{a_2-a_1}} \cdot10^{(a1-a3)\cdot t}\\
    &=\frac{M_*}{1.4\times10^{8}}\cdot10^{-0.113\cdot t}.
\end{align*}
It follows that the sSFR for main sequence galaxies with low stellar masses (well below $\mathrm{M}_\mathrm{T}(t)$) does not effectively depend on their concrete stellar mass, but instead only on cosmic time $\mathrm{sSFR}\,[1/\mathrm{yr}]=\mathrm{SFR}/M_*=7\times10^{-9}\cdot10^{-0.113\cdot t}$. This implies that the steady state for star formation in low mass galaxies ($M_*<10^{9}\mathrm{M}_\odot$) becomes increasingly inefficient toward lower redshifts (matching findings e.g. \citealp{ilbert:2015}), even though some studies suggest there to be little evolution in the total gas content of the galaxies \citep{merida:2024}. Furthermore, observations of high-redshift galaxies are showing elevated mass loading factors, implying stronger outflows that could more effectively suppress their star formation relative to their low redshift counterparts \citep{carniani:2024}. Similarly, the metallicity of the gas in these galaxies increases toward lower times \citep{maiolino:2008,curti:2024}, which would imply faster, more efficient cooling \citep{wiersma:2009}. Instead, this may be due to an increase in the angular momentum of the infalling gas (seen for more massive galaxies in cosmological simulations by \citealp{teklu:2023} and \citealp{fortune:2025}), which would depress the central gas density and thus SFR, while some theoretical works find feedback to be less efficient at low metallicity \citep{jecmen:2023}.

\begin{figure}
\centering
\includegraphics[width=0.9\columnwidth]{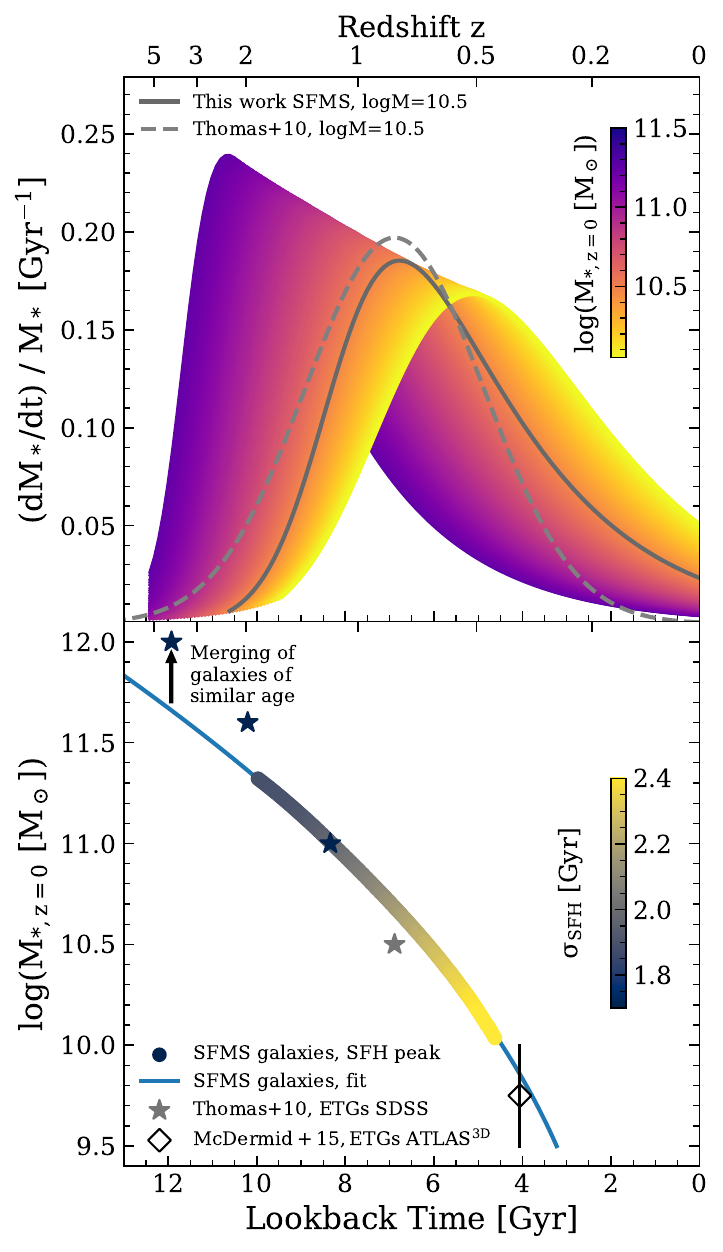}
\caption{\textbf{Top:} The specific star formation rate history of massive galaxies that fully evolved along the SFMS, colored by their final stellar mass $\mathrm{M}_\mathrm{*,z=0}$. Plotted are only models where the progenitor galaxy accounts for less than $1\%$ of $\mathrm{M}_\mathrm{*,z=0}$. \textbf{Bottom:} The peak location of best-fit Gaussians to each SFH versus $\mathrm{M}_\mathrm{*,z=0}$, colored by the width of the Gaussian. From observations of early-type galaxies we include the fits of \citet{thomas:2010} for SDSS galaxies, and the peak location of the SFH for lower mass galaxies in ATLAS$^\mathrm{3D}$ by \citet{mcdermid:2015}. We do not include peak locations for the higher mass bins by \citet{mcdermid:2015} as they peak at $z=10$.}\label{fig:f_age}
\end{figure}

For large stellar masses $M_*\gg \mathrm{M_\mathrm{T}(t)}$ the fraction $\mathrm{M}_\mathrm{T}(t)/M_*$ tends toward zero, such that 
\begin{align*}
    \mathrm{SFR}(\mathrm{high\, M}_*,t)&\approx \mathrm{SFR}_\mathrm{max}(t)= 10^{a_0+a_1t}\\
    &=512.9 \,\cdot\, 10^{-0.186t},
\end{align*}
which means that, unlike low mass galaxies, massive galaxies retain a $1/M_*$ dependence for their sSFR. However, as their sSFR is so low and thus their stellar mass does not noticeably increase at later times (see the top panel of Fig.~\ref{fig:intro}), effectively their sSFR again mostly depends on the cosmic time.

Finally, we see that more massive galaxies peak in their SFR earlier in time and at higher values, and in particular that they form the majority of their stars faster than lower mass galaxies. To quantify this, we plot in the top panel of Fig.~\ref{fig:f_age} the formed stellar mass as a function of lookback time, normalized by the total formed stellar mass (note that this is not equal to the final, i.e., remaining stellar mass). Here we seed progenitors with stellar masses between $1\times10^{8}$ and $1\times10^{9}\mathrm{M}_\odot$ and redshift bins from $z=0.5$ to $z=6$, but keep only those models where the progenitor mass never accounts for more than $1\%$~of the final stellar mass (i.e., the evolution on the SFMS dominates the stellar ages).

We see from the top panel of Fig.~\ref{fig:f_age} that for purely main sequence-evolved galaxies, there naturally emerges a more peaked, early star formation history (SFH) in more massive galaxies, while lower mass galaxies have broader distributions and peak later toward $t=6\,\mathrm{Gyr}$. To quantify the behavior more clearly, we fit a Gaussian function of the form $y = 1/(\sqrt{2\pi}\sigma_\mathrm{SFH})\cdot\exp^{-((x-\mu)/\sqrt{2}\sigma_\mathrm{SFH})^2}$ to the SFHs, and plot the location of the peaks $\mu$ (in lookback time) against the final stellar mass in the bottom panel of Fig.~\ref{fig:f_age}. We find a very tight relation that we fit (blue line) as $m = 0.9408\cdot T^2 -15.88\cdot T+69.18$, with $T \equiv \mu_\mathrm{peak}\, [\mathrm{Gyr}]$ and $m=\log(\mathrm{M}_{*,z=0}\,[\mathrm{M}_\odot])$.

The behavior found for SFMS evolved galaxies is very similar to the distribution of stellar ages found by \citet{thomas:2010} when modeling early-type galaxies (colored stars), where those with $\mathrm{M}_{*,z=0}=10^{10.5}\mathrm{M}_\odot$ peak at $T\approx6.9\,\mathrm{Gyr}$ compared to our later $T\approx6.2\,\mathrm{Gyr}$. This is likely due to quenching, as model SFMS galaxies of $\mathrm{M}_{*,z=0}=10^{10.7}\mathrm{M}_\odot$ have the same peak location such that, if we cut their later evolution short through quenching, they would end up at lower mass where \citet{thomas:2010} find their ETGs. This agreement persists also for galaxies of $\mathrm{M}_{*,z=0}=10^{11}\mathrm{M}_\odot$, which have identical peak ages of $T\approx8.3\,\mathrm{Gyr}$ for both our model SFMS galaxies and their SDSS ETGs.

For those of $\mathrm{M}_{*,z=0}>10^{11}\mathrm{M}_\odot$, forming that many stars along the SFMS becomes increasingly difficult (due to the flattening in SFR at high $M_*$) and requires starting well before $z>6$ where the fit by \citealp{popesso:2023} does not apply any longer. This simply indicates that such galaxies must be formed either through particularly strong starbursts (potentially driven by wet mergers at high redshifts), or more likely through the merging of multiple old, massive galaxies at later times. Such merging increases the mass without altering the peak ages, as indicated by the arrow in the bottom panel of Fig.~\ref{fig:f_age}. Nevertheless, the fact that the model predicts a very similar relation to observations even of ETGs highlights the self-similarity of the galaxy growth for a given stellar mass until the moment of dry merging and quenching.

\begin{figure}
\centering
\includegraphics[width=0.9\columnwidth]{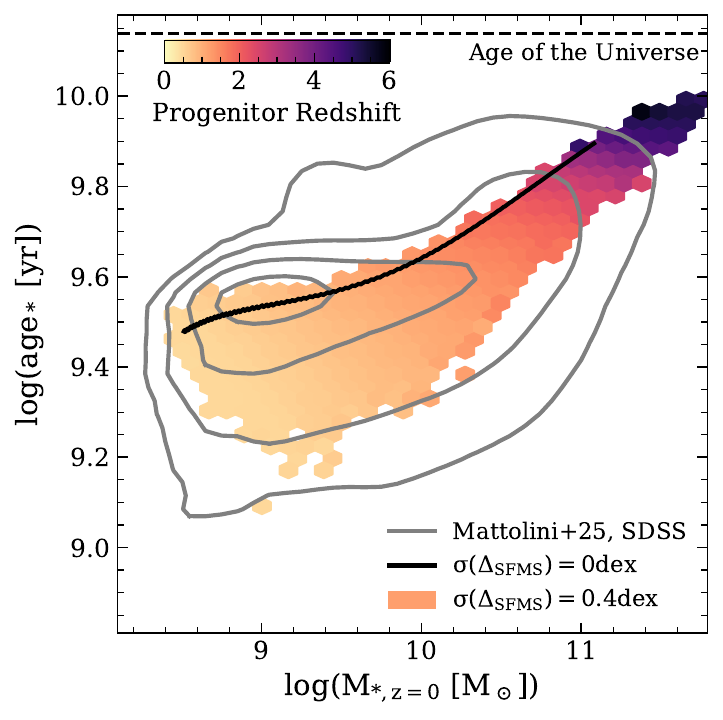}
\caption{Mass weighted mean stellar age plotted versus stellar mass at $z=0$. The black solid line shows the ages for galaxies evolving exactly along the SFMS, while the heatmap is the distribution when including a lognormal scatter in SFR with $\sigma=0.4$dex. Contour lines are from observed SDSS galaxies by \citet{mattolini:2025}. Interestingly, the observed ages very closely follow the expectations from the SFMS evolution, in particular when we allow for some scatter about the main sequence.}\label{fig:mstr_age}
\end{figure}

We then consider the remaining stellar mass at $z=0$ (instead of the total formed mass through time), and determine the mass-weighted ages as a function of the total $\mathrm{M}_{*,z=0}$. From this we obtain the solid black line in Fig.~\ref{fig:mstr_age}, which follows the age distribution of a large sample of SDSS galaxies by \citet{mattolini:2025} as indicated by the orange contour lines remarkably well. However, the requirement that our model galaxies formed their stars exactly along the SFMS necessarily leads to a lack of scatter, where there is a noticeable range in ages present in the observed galaxies. To more closely mirror the observed galaxies, we therefore allow for a lognormal scatter around the SFMS with a standard deviation of $0.4\,$dex to obtain the heatmap (with the color being the redshift of the progenitor galaxy). We find that this fills out the range in contours by \citet{mattolini:2025} well. Naturally, the extreme cases of very young (recent starbursts) and very old (quenched and/or dry merged galaxies) are not captured, but the overall behavior is. Consequently, we conclude that for the majority of stars in present-day galaxies, their age distribution is essentially as would be expected if the host had evolved with some scatter around the SFMS. 

\begin{figure*}
\centering
\includegraphics[width=0.9\textwidth]{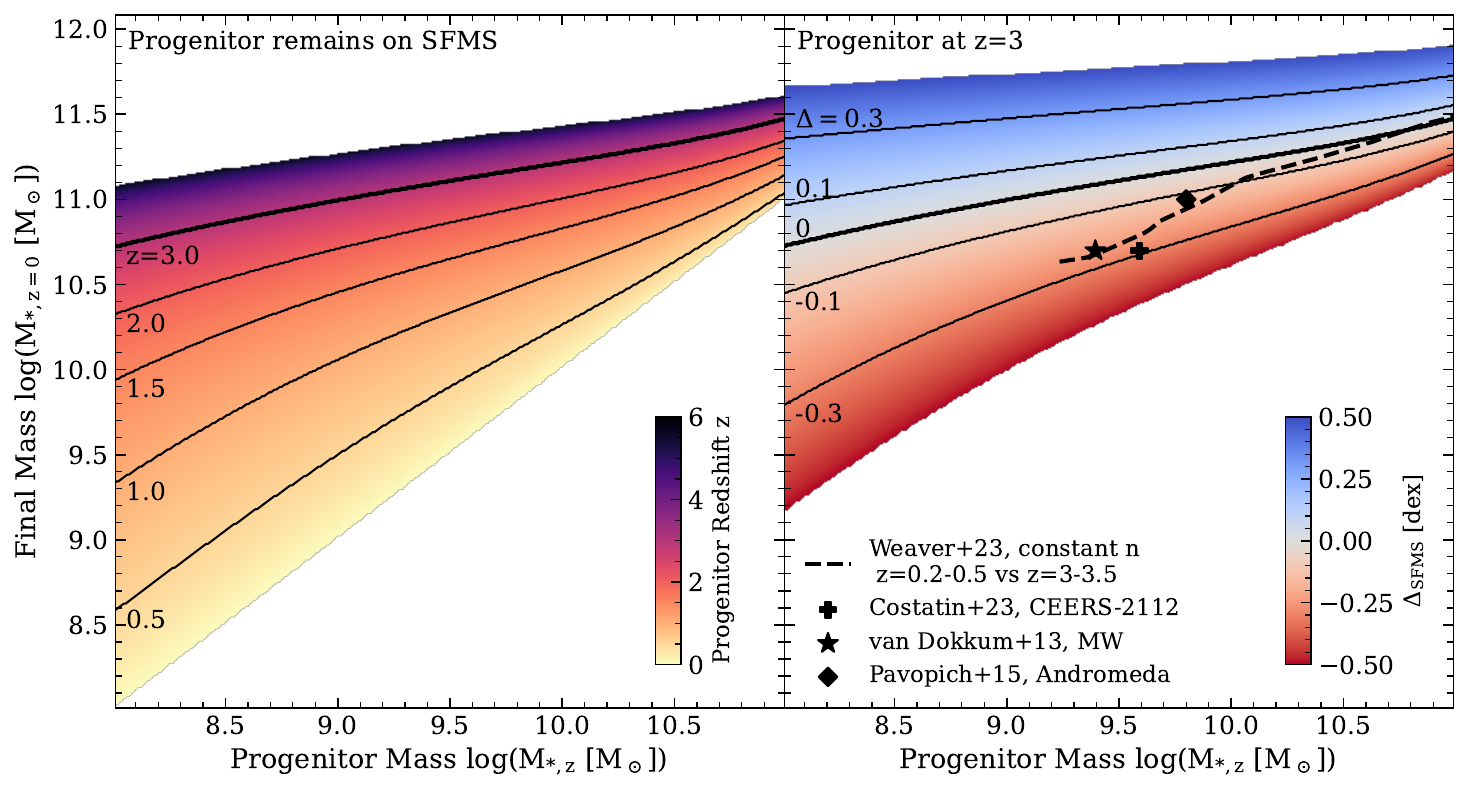}
\caption{The final stellar mass of a given galaxy plotted against its progenitor mass. On the \textbf{left} we vary the redshifts of the progenitor galaxies (given by the color) and then evolve them precisely on the SFMS. On the \textbf{right} the progenitors are fixed at $z=3$ while the SFR of the galaxies may lie consistently above (blue) or below (red) the expected SFR from the SFMS. The thick solid black line is identical for both panels (SFMS evolved galaxies with progenitors at $z=3$). In the right panel, we additionally plot progenitor stellar masses of the MW and Andromeda galaxies from \citet{vandokkum:2013} and \citet{papovich:2015}, as well as the possible MW progenitor CEERS-2122 by \citet{costantin:2023}. It follows that the MW and Andromeda must have either undergone cycles of bursty and then suppressed star formation, or instead formed stars since $z=3$ at a rate consistently below the SFMS by 0.3 and $0.1\,$dex, respectively. We also plot the relation between progenitor and descendant galaxy under the assumption that they lie at constant number density (black dashed line) using the \citet{weaver:2023} stellar mass function.}\label{fig:grid_chab}
\end{figure*}

It is interesting to note that the ages of our model galaxies mirror the observational trends even for the more massive galaxies that are predicted to grow primarily through accretion at later times \citep{oser:2010,remus:2023}, which we do not include in the model at all. The good agreement in the ages implies that the amount of younger stars brought in during these merging events is comparable to the amount the galaxy would have formed in-situ if the baryons had instead been accreted as gas. In that case, the galaxy could have remained on the SFMS down to $z = 0$. This is likely because young stars in massive galaxies contribute little to the overall mass budget. As a result, differences between late in-situ star formation and merging activity have no discernible effect on the mass-weighted average age, although they may affect luminosity-weighted ages.

\subsection{Connecting galaxies across cosmic time}\label{subsec:connect}

We then consider the relation between the progenitor galaxies and the final stellar mass. This relation is crucial when connecting high-redshift and low-redshift galaxies, to understand what type of galaxies at high redshifts are possible progenitors of present-day main sequence galaxies. In addition, it informs us which stellar masses would be expected if the observed massive star-forming galaxies at high redshifts \citep[e.g.,][]{wang:2025} remained on the SFMS down to $z=0$, and if they would grow too massive.

To this end, we construct a grid of models with progenitor masses between $M_*=1\times10^{8}\mathrm{M}_\odot$ and $M_*=1\times10^{11}\mathrm{M}_\odot$, with redshifts from $z=0$ to $z=6$, and plot their final $z=0$ stellar mass assuming they evolve entirely along the \citet{popesso:2023} SFMS in the left panel of Fig.~\ref{fig:grid_chab}. The colorbar indicates the redshift of the progenitor, with the solid black lines highlighting particular redshifts. We see that varying the mass of the progenitor galaxy does not strongly impact the resulting final mass if it is seeded at high redshifts (purple region), because in all cases the galaxies reach the ``turnaround mass'' before $z=0$, above which the sSFR becomes too low to significantly grow the galaxies any further (as discussed also for Fig.~\ref{fig:intro}). 

We find that even a low mass progenitor of $M_*=1\times10^{8}\mathrm{M}_\odot$ reaches $\mathrm{M}_{*,z=0}=1\times10^{11}\mathrm{M}_\odot$ by $z=0$ if it evolves along the SFMS from $z=6$ onward. Conversely, because the high redshift lines lie at very high final stellar masses for the whole range of progenitor masses, galaxies today with $\mathrm{M}_{*,z=0}=10^{10}\mathrm{M}_\odot$ or lower either had their progenitor emerge particularly late in cosmic time, past the peak of the cosmic star formation rate density \citep{madau:2014} at $z=1.5$ even for a dwarf of $M_*=1\times10^{8}\mathrm{M}_\odot$, or they must have undergone phases of suppressed star formation or even quiescence. 
Two interesting examples for this can be found directly in the local Universe: M33, the third most massive galaxy in our local group, has a stellar mass of $4.8\times10^{9}\mathrm{M}_\odot$ \citep{corbelli:2014}. With respect to many of its properties, NCG~300 with a stellar mass of $2\times10^{9}\mathrm{M}_\odot$ \citep{munoz:2007} is often treated as its analogue, with the difference that NGC~300 lives within a cosmic void. Both galaxies are close enough that stars can be resolved observationally. Based on such a study, \citet{gogarten:2010} find that NCG~300 had formed about 70\% of its stars, that is $1.4\times10^{9}\mathrm{M}_\odot$, already at $z=2$, while M33 has formed only about 20-30\% at $z\approx2$, i.e., $1.4\times10^{9}\mathrm{M}_\odot$. While M33 after $z=2$ can have continued to form stars on the main sequence, albeit with an SFR continuously about -0.5dex below the main sequence, NGC~300 must have gone through a period of relative quiescence with SFR more than -0.5dex below the SFMS (see Fig.~\ref{figA:grid_chab} in the Appendix for the evolutionary tracks). Indeed for M33, \citet{williams:2009} find indications for a rather low-star formation period 8-10$\,$Gyrs ago, with the more recent star formation taking place mostly in the outskirts. This clearly demonstrates that significant amounts of old stars detected in low mass galaxies is a smoking gun for a period of quiescence in the galaxy's lifetime.

Furthermore, cycles of rejuvenation and quiescence are indeed found from cosmological simulations both at later times \citep{fortune:2025} as well as already at high redshifts \citep{remus:2025}, with indications for rejuvenation also seen from observations \citep[e.g.,][]{peletier:2007,munozlopez:2025}. Any amount of ex-situ stars (i.e., mergers), which we do not consider here, would only exacerbate the need for such suppressed phases by decreasing the amount of time allowed on or near the SFMS. 

For the case of super spirals with $\mathrm{M}_{*,z=0}\approx10^{12}\mathrm{M}_\odot$ \citep{ogle:2019}, we find there to be no way to produce their stellar mass in-situ while evolving entirely on the SFMS, even for extreme progenitors of $M_*=10^{11}\mathrm{M}_\odot$ at $z=6$, such as the disk galaxy reported by \citet{xiao:2025}. This implies that they must have either undergone significant amounts of merging or evolved consistently above the main sequence. To test the second possibility, we construct a grid of progenitors but fix their redshift at $z=3$, and instead allow their SFR to evolve at a fixed offset $\Delta_\mathrm{SFMS}$ from the SFMS going from $-0.5\,$dex (red) to $+0.5\,$dex (blue) in the right panel of Fig.~\ref{fig:grid_chab}. The solid black lines denote particular values of $\Delta_\mathrm{SFMS}$, while the thick solid black line is the same in both panels (SFMS evolved galaxies seeded at $z=3$).

We find for the case of $\Delta_\mathrm{SFMS}=0.5\,$dex, meaning the galaxy has consistently been above the SFMS since $z=3$, that in-situ formation of enough stars to match the observed super spirals by \citet{ogle:2019} is possible, though it is doubtful whether galaxies can remain so far above the SFMS over the majority of the age of the Universe. Given their largely undisturbed morphologies in observations, this points toward very significant contributions from (gas-rich) minor mergers or accretion directly from the cosmic web that would leave the disk structure intact \citep{karademir:2019}, or progenitors in excess of $M_*=10^{11}\mathrm{M}_\odot$ by $z=3$.

Generally, for galaxies with progenitors $M_*>10^{8}\mathrm{M}_\odot$ present by $z=3$, we find that they must either end up being very massive at $z=0$ or have spent a majority of their time with SFRs well below the SFMS mean. In particular, we highlight here the cases of Milky Way- and Andromeda-mass galaxies. Under the assumption that the present day number density of galaxies of a given stellar mass is comparable to the number density of their progenitors, \citet{vandokkum:2013} and \citet{papovich:2015} determine the expected progenitor galaxy stellar masses at $z=3$ of the Milky Way and Andromeda galaxies to be $M_*\approx2.5\times10^9\mathrm{M}_\odot$ and $M_*\approx6.3\times10^9\mathrm{M}_\odot$, respectively. We plot those points, as well as the potential Milky Way progenitor CEERS-2112 with $M_*\approx3.9\times10^9\mathrm{M}_\odot$ from \citet{costantin:2023}, in the right panel of Fig.~\ref{fig:grid_chab}. 

Fig.~\ref{fig:grid_chab} shows that Milky Way-mass galaxies must have formed stars at a rate nearly $0.3\,$dex below the SFMS value since $z=3$ if they are to end up with their current day stellar mass of $\mathrm{M}_{*,z=0}\approx5\times10^{10}\mathrm{M}_\odot$, and that is under the assumption of no additional merging. Indeed, there are well-known indications from the ages of stars in the Milky Way itself that indicate a low star formation rate after $z<1$ \citep{haywood:2016}. In contrast, the Andromeda galaxy at $\mathrm{M}_{*,z=0}\approx1\times10^{11}\mathrm{M}_\odot$ is massive enough that, while still suppressed, it could have evolved from $z=3$ progenitors of the same number density with an SFR just $0.1\,$dex below the SFMS value. 

Going one step further, we have applied a similar number density methodology to the more recent census of the galaxy stellar mass function from $z=6$ to $z=0.1$ on the COSMOS2020 catalog done by \citet{weaver:2023}. We interpolate between their data points in log-log space to produce the smooth black dashed line in the right panel of Fig.~\ref{fig:grid_chab}, which shows the relation between $z=3$ and $z=0$ galaxy stellar masses of constant number density. Naturally, the values by \citet{vandokkum:2013} for the Milky Way and \citet{papovich:2015} for Andromeda lie near this line, while the potential Milky Way progenitor from \citet{costantin:2023} has a somewhat higher stellar mass. We find, if the number density connection is correct, that galaxies of $\mathrm{M}_{*,z=0}=3\times10^{10}\mathrm{M}_\odot$ must have spent the majority of their evolution below the SFMS, while, curiously, more massive galaxies of $\mathrm{M}_{*,z=0}>2\times10^{11}\mathrm{M}_\odot$ could have remained entirely on the SFMS ($\Delta_\mathrm{SFMS}=0$) down to $z=0$. As those more massive ones are instead expected to grow primarily through mergers at later times, the accreted stellar mass must approximately equal the expected in-situ formed amount if they retained an SFR on the SFMS. 

Fig.~\ref{fig:grid_chab} is particularly useful for constraining possible star formation histories of galaxies evolving near the SFMS over cosmic time, and connecting possible past progenitors with present-day galaxies. As it is limited by the choice of progenitor redshift, we include six versions with progenitors at redshifts $z=6$ to $z=1$ in the Appendix~\ref{app:finding}.

\subsection{Milky Way-like Galaxies}\label{subsec:mw}

\begin{figure}
\centering
\includegraphics[width=0.9\columnwidth]{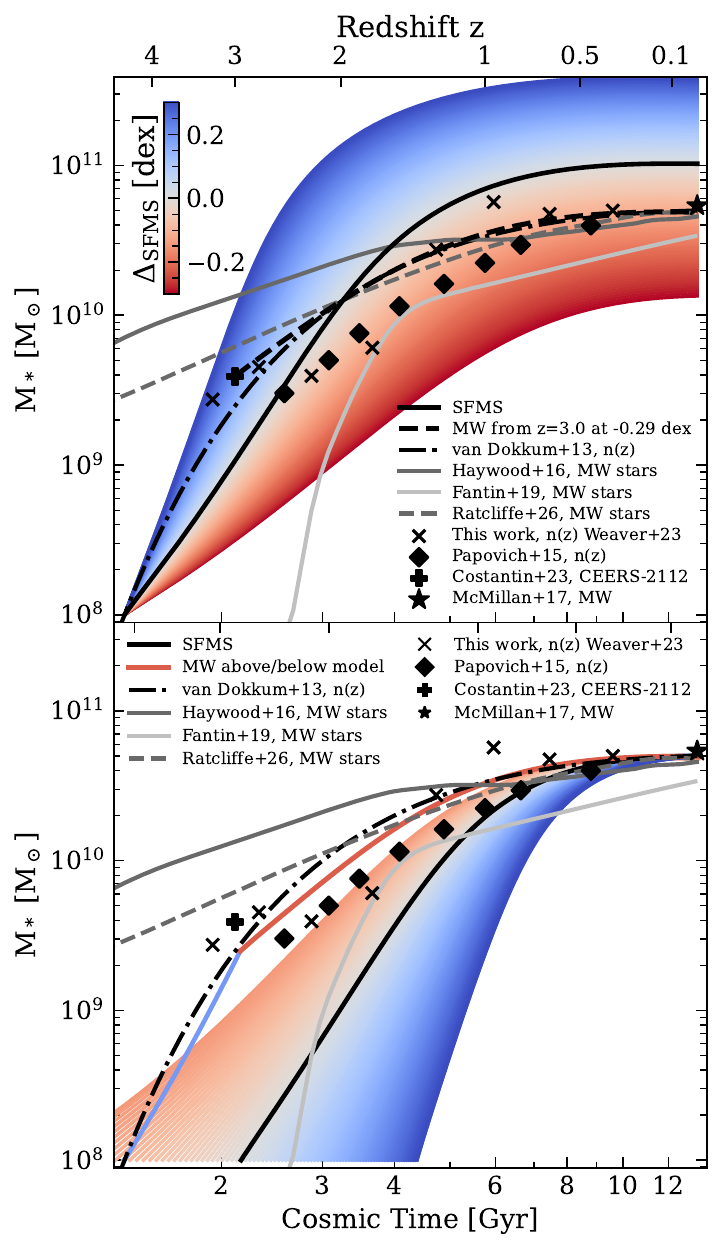}
\caption{The evolution of the stellar mass of Milky Way-like galaxies that formed stars along (solid black line), above (blue) or below (red) the SFMS, plotted against time. The \textbf{top} panel shows expectations for a progenitor galaxy with $M_*=1\times10^{8}\mathrm{M}_\odot$ at $z\approx4.4$, following the prediction from \citet{vandokkum:2013} made on the basis of number density arguments (dash-dotted black line). Also plotted are the data by \citet{papovich:2015} obtained in a similar manner (diamonds), our own version obtained from the \citet{weaver:2023} COSMOS2020 data (crosses), the integrated stellar mass from the SFRs (solid and dashed gray lines) by \citet{haywood:2016} and \citet{ratcliffe:2025}, the cumulative mass (light gray) by \citet{fantin:2019}, the potential MW progenitor CEERS-2112 (plus sign) by \citet{costantin:2023}, and the $z=0$ MW mass (star) determined by \citet{mcmillan:2017}. The \textbf{bottom} panel shows instead the possible galaxy stellar mass tracks that obtain a MW-like stellar mass at $z=0$ $\mathrm{M}_{*,0}=5\times10^{10}\mathrm{M}_\odot$, and a model that closely resembles the \citet{vandokkum:2013} predictions (two-colored solid line).}\label{fig:mw_evo}
\end{figure}

Having seen that galaxies of Milky Way mass ($M_{*,z=0}=5\times10^{10}\mathrm{M}_\odot$) appear to require a suppressed late time SFR to match expectations from number density arguments, it is worthwhile to consider a more detailed look at model star formation histories. We show in Fig.~\ref{fig:mw_evo} the stellar mass against cosmic time, in the top panel for a $M_{*}=1\times10^{8}\mathrm{M}_\odot$ progenitor galaxy set at $z=4.4$ based on the \citet{vandokkum:2013} fit for the Milky Way's evolution, and in the bottom panel for a set of model galaxies that all end at $\mathrm{M}_{*,z=0}=5\times10^{10}\mathrm{M}_\odot$.

Beginning with the top panel of Fig.~\ref{fig:mw_evo}, we find that the progenitor would noticeably overshoot the mass of the MW if it evolved along the SFMS, ending with $\mathrm{M}_{*,z=0}=1\times10^{11}\mathrm{M}_\odot$. This is particularly interesting, as the early growth of the \citet{vandokkum:2013} evolution (dash-dotted black line) actually exceeds the expected SFMS evolution (solid black line) of the progenitor by around $+0.3\,$dex until $z=3$, when it nears the stellar mass of the possible MW progenitor observed by \citet{costantin:2023}. At that point, the growth slows significantly relative to the SFMS curve, very closely matching a suppressed SFR with $\Delta_\mathrm{SFMS}=-0.3\,$dex (dashed black line) until $z=0$. The same holds true for the predictions (diamonds) by \citet{papovich:2015}, which lie near the SFMS line at $z=2$ but afterwards deviate toward negative $\Delta_\mathrm{SFMS}$ values.

We then derive our own updated evolutionary track (crosses) from the COSMOS2020 galaxy stellar mass function by \citet{weaver:2023}. To do so, we first determine the number density of Milky Way mass galaxies ($\mathrm{M}_{*,z=0}=5\times10^{10}\mathrm{M}_\odot$) at $z=0.2$ to $z=0.5$ as $n_\mathrm{MW}=2.2\times10^{-3}\mathrm{cMpc^{-3}}$, and then find the stellar mass that matches this number density at higher redshifts by linearly interpolating in log-log space between the \citet{weaver:2023} data points. Furthermore, we fit an orthogonal distance regression to these new data points with the same function form as that by \citet{vandokkum:2013}, and obtain for the galaxy stellar mass evolution of Milky Way-mass galaxies:
\begin{equation}\label{eq:mw_evo}
    \mathrm{M}_{*,\mathrm{MW}}(z) = 10^{10.74-0.007\cdot z-0.17\cdot z^2}.
\end{equation}
This fit is largely similar to that determined by \citet{vandokkum:2013}, though with a steeper decrease in stellar mass at higher redshifts. 

It is further striking that our integration of the $\mathrm{SFR}(t)$ determined by \citet{haywood:2016}, which they derived from chemical abundances of Milky Way stars following \citet{snaith:2015}, results in a particularly early star formation history, with a stellar mass of $M_{*}=1\times10^{10}\mathrm{M}_\odot$ by $z=3.6$. We note that, to obtain a final mass $M_{*,z=0}\approx5\times10^{10}\mathrm{M}_\odot$ it was necessary to assume a steep Chabrier IMF, which is more akin to the \citet{kroupa:1993} IMF. If we instead assume a normal Chabrier as done for our SFMS models (and which is also closer to the \citealp{kroupa:2001} IMF used by \citealp{haywood:2016}), we obtain a final stellar mass of $M_{*,z=0}\approx3.6\times10^{10}\mathrm{M}_\odot$. While this does reduce the early stellar mass by the same fraction, a stellar mass of $M_{*}=1\times10^{10}\mathrm{M}_\odot$ is still reached much earlier (by $z=3.1$) than found by the curves from number density arguments. Newer data by \citet{ratcliffe:2025} using giant stars from APOGEE finds a similarly early formation of the Milky Way, while the history by \citet{fantin:2019} using white dwarf stars rises significantly later. Note that for the latter we use their SFH directly, as opposed to integrating their given SFR. 

\begin{figure}
\centering
\includegraphics[width=0.95\columnwidth]{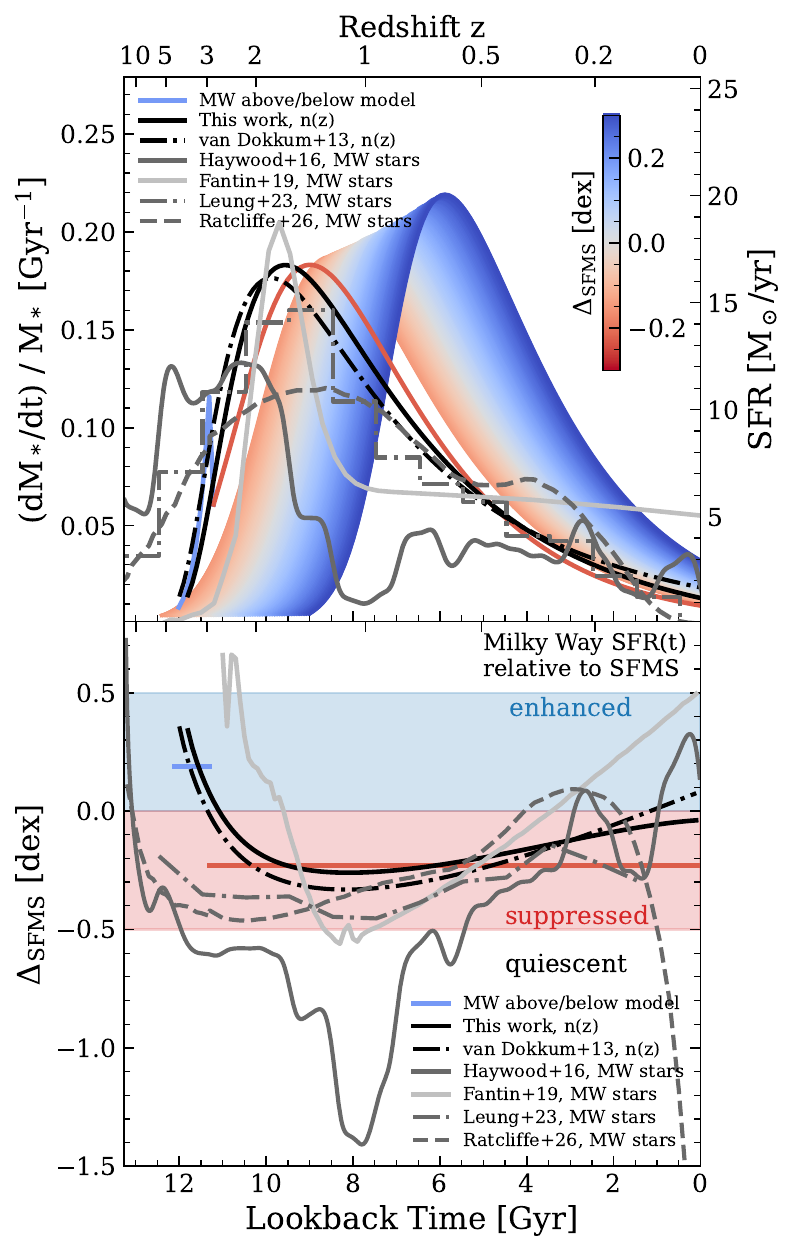}
\caption{The normalized star formation histories of galaxies with a Milky Way-like $z=0$ stellar mass $\mathrm{M}_{*,0}=5\times10^{10}\mathrm{M}_\odot$ that evolved consistently on (solid black line), above (blue) or below (red) the SFMS. Because the total mass for all curves is the same, the y-axis translates directly to an SFR. We show also the SFH derived from \citet{vandokkum:2013}, the age distribution by \citet{leung:2023} and SFR by \citet{ratcliffe:2025} based on APOGEE data of Milky Way giant stars, the SFH by \citet{fantin:2019} recovered from white dwarfs, and the SFR against time derived by \citet{haywood:2016} through the alpha abundances of Milky Way stars via the method from \citet{snaith:2015}. Note that the data by \citet{fantin:2019} is explicitly plotted as SFR and does not integrate to the same final mass (see Fig.~\ref{fig:mw_evo}).}\label{fig:mw_f_age}
\end{figure}

For the bottom panel of Fig.~\ref{fig:mw_evo}, we see that if the Milky Way had evolved entirely on the SFMS, its progenitor would fall below a stellar mass of $\mathrm{M}_{*}=1\times10^{8}\mathrm{M}_\odot$ above $z=2.8$. Given the observational evidence for significant amounts of older stars from directly within the Milky Way itself \citep{haywood:2016,imig:2023,leung:2023} this is clearly not the case. Consequently, our Milky Way galaxy must have followed an SFH that sees an early, bursty rise in stellar mass, after which its SFR was strongly suppressed well below the SFMS. Indeed, this is the scenario derived by \citet{haywood:2016} from the $\alpha$-abundances of stars in the Milky Way. We find that a model with $\Delta_\mathrm{SFMS}=0.189\,$dex for $z>3$ and $\Delta_\mathrm{SFMS}=-0.23\,$dex after provides a good representation of the evolution by \citet{vandokkum:2013}.

From the mass evolution, we can derive also the resulting SFH, shown in the top panel of Fig.~\ref{fig:mw_f_age}, which can be directly converted to the SFR because the final stellar mass $\mathrm{M}_{*,z=0}=5\times10^{10}\mathrm{M}_\odot$ is approximately the same for all curves. By necessity, models which grow later but with higher SFRs (positive $\Delta_\mathrm{SFMS}$, blue) show more peaked SFHs, while allowing for a more suppressed SFR (negative $\Delta_\mathrm{SFMS}$, red) broadens the SFH but crucially also shifts the peak toward earlier times. This better matches expectations for the Milky Way itself, as can be seen from the SFH by \citet{vandokkum:2013}, as well as the age distribution and SFR derived from Milky Way stars directly by \citet{haywood:2016} and \citet{leung:2023}, respectively. Note that for the \citet{vandokkum:2013} curve we have converted their fit of the observed stellar mass $M_*(t)$ (equal to the formed minus lost mass) into an $\mathrm{SFR}(t)$ (which is just the formed mass) by splitting the evolution into timesteps and accounting for stellar mass loss following a Chabrier IMF.

Finally, the combination of $\mathrm{SFR}(t)$ and $M_*(t)$ allows us to directly compare the Milky Way SFR with the expectation SFR from the \citet{popesso:2023} SFMS through cosmic time, which we plot in the bottom panel of Fig.~\ref{fig:mw_f_age}. All methods find that the Milky Way begins its star formation particularly early and bursty, lying around $0.3\,$dex above the SFMS for the number density based models and $0.5\,$dex for the \citet{haywood:2016} data. We note that for both cases this occurs at a stellar mass within the bounds of the \citet{popesso:2023} main sequence fit $M_*>3\times10^8\mathrm{M}_\odot$, but for the latter at earlier redshifts than their fits. It is thus possible that the extrapolation is incorrect here, and the early Milky Way could still have evolved along the SFMS. Then, between a redshift of $z=3$ and $z=2$, all models find the Milky Way SFR to decrease relative to the SFMS, dropping down to $\Delta_\mathrm{SFMS}=-0.4\,$dex for the number density models and even farther below for those based on Milky Way stars. The SFR remains suppressed until very late times, when in the last gigayear it lies on the main sequence again.

We note that this rise toward the SFMS at later times is actually driven entirely by the SFMS itself coming down, as can be seen from the absolute SFR in the top panel of Fig.~\ref{fig:mw_f_age}. The SFRs at later times of our Milky Way galaxy are flat \citep{haywood:2016} or decreasing \citep{vandokkum:2013,leung:2023}, but the SFMS sinks rapidly enough that they still meet by $z=0$. This matches the behavior of the ``caught'' type of galaxies defined by \citet{fortune:2025} within cosmological simulations, which exhibit also other properties that are similar to those of the Milky Way. They are found there to be more common than galaxies that evolved entirely on the SFMS, though the most common type are rejuvenating galaxies that go through cycles of SF and quiescence on timescales $t>1\,$Gyr (and indeed, \citealp{ruiz_lara:2020} find indications within the Milky Way itself that Sagittarius may have triggered cycles of enhanced star formation). For the case of the \citet{haywood:2016} SFH, the Milky Way galaxy would actually fall under this latter category. With a more stringent SFMS cutoff at $\pm0.3\,$dex, essentially all our considered SFHs of the Milky Way would classify it as a ``rejuvenated'' galaxy \citep{fortune:2025}.

\section{Conclusion}\label{sec:conc}

We have developed a model that allows us to trace the evolution of galaxy stellar masses and star formation rates (SFR) through cosmic time in relation to their position on the star-forming main sequence of galaxies (SFMS) described by \citet{popesso:2023}. With this model, we explored the star formation histories and stellar mass growth of galaxies of differing masses, their resulting age distributions and in particular highlighted the example of the Milky Way galaxy. 

We found that the observed difference in age distributions between galaxies of lower and higher stellar masses naturally arises from the evolution of the star-forming main sequence through cosmic time (Fig.~\ref{fig:f_age}-\ref{fig:mstr_age}). The model predicts that low-mass galaxies are younger with extended star formation histories, whereas massive galaxies form earlier and more rapidly \citep{thomas:2010,mattolini:2025}. Galaxies that evolved entirely along the star-forming main sequence match the observed age distributions for galaxies from $M_*=1\times10^{9}\mathrm{M}_\odot$ up to $M_*=3\times10^{11}\mathrm{M}_\odot$ remarkably well, with only a slight tendency toward younger ages due to the lack of quenching. 

Based on our model, we provided look-up figures relating present-day stellar masses with possible progenitors at redshifts from $z=1$ up to $z=6$, accounting also for scatter above/below the star-forming main sequence of up to $\pm0.5\,$dex. These figures can be used to inform current observations of high redshift galaxies about the expected stellar masses should they continue to evolve along, above or below the star-forming main sequence down to $z=0$. 

We showed that present-day high-mass galaxies ($M_*\geq1\times10^{11}\,\mathrm{M}_\odot$) grew their stellar mass as expected if they had evolved along the star-forming main sequence, based on the COSMOS2020 galaxy stellar mass function \citep{weaver:2023} and under the assumption that galaxies at fixed number density are connected through cosmic time. More massive galaxies are expected to grow secularly \citep{oser:2010}. Our results therefore suggest that their total stellar mass growth is largely independent of whether baryons are accreted as gas and form stars in situ, or are added directly as stars through mergers.

We applied our model to the Andromeda and Milky Way galaxies. As Andromeda is more massive than the Milky Way ($M_*\approx1\times10^{11}\mathrm{M}_\odot$ versus $M_*\approx5\times10^{10}\mathrm{M}_\odot$) we find that the predicted growth history by \citet{papovich:2015} matches the expectations from an evolution along the SFMS (lying on average at $\Delta_\mathrm{SFMS}=-0.1\,$dex). By contrast, we show that galaxies which end up with a present-day Milky Way stellar mass ($M_*\approx5\times10^{10}\mathrm{M}_\odot$) must have evolved with star formation rates well below the SFMS ($\Delta_\mathrm{SFMS}\approx-0.3\,$dex) since $z=3$, if they are to remain at the same comoving number density through cosmic time. Any additional contributions from mergers only further reduce the allowed SFR. The lowest point in SFR relative to the SFMS of Milky Way-mass galaxy progenitors is found to be around $z=1$, after which the distance shrinks and their SFRs approach the SFMS again. This behavior is inverted before $z=3$, meaning that to remain at a constant number density, the progenitors of Milky Way-mass galaxies would have had to evolve significantly \emph{above} the SFMS. We provide a fit for the growth of Milky Way-mass galaxies as $\log\mathrm{M}_{*,\mathrm{MW}}(z) = {10.74-0.007\cdot z-0.17\cdot z^2}$, based on the data by \citet{weaver:2023}.

For the Milky Way, we demonstrate that quenching ($\Delta_\mathrm{SFMS}\leq-0.5\,$dex) at $z=1$ is implied by chemical abundance data from giant stars \citep{haywood:2016,leung:2023,ratcliffe:2025} and white dwarf stars \citep{fantin:2019}. This drop in star formation rate is stronger than expected from constant number density arguments, meaning that the Milky Way grew earlier and underwent a more significant period of suppressed star formation than other galaxies of comparable mass. We conclude that rejuvenated star formation is increasingly observed in other galaxies, both in observations \citep{munozlopez:2025} and simulations \citep{fortune:2025,remus:2025}. This scenario warrants further exploration, in particular with regard to the assumed shapes of star formation histories used in SED fitting codes \citep{carnall:2018}.

\begin{acknowledgements}
LCK acknowledges support by the Deutsche Forschungsgemeinschaft (DFG, German Research Foundation) under project nr. 516355818 and the COMPLEX project from the European Research Council (ERC) under the European Union’s Horizon 2020 research and innovation program grant agreement ERC-2019-AdG 882679. LCK and RSR acknowledge support by DFG under Germany's Excellence Strategy -- EXC-2096 -- 3900783311. RLD is supported by the Australian Research Council through the Discovery Early Career Researcher Award (DECRA) Fellowship DE240100136 funded by the Australian Government.

The following software was used for this work: Julia \citep{bezanson+17:julia}.
\end{acknowledgements}

\bibliographystyle{aa}
\bibliography{mw_growth}

\begin{appendix}

\section{Impact of the model parameters}
\label{app:timesteps}

Throughout this work, we considered one setup of a main sequence evolution model with a timestep size of the iteration process of $\delta t=100\,\mathrm{Myr}$. Therefore, it is worth discussing how alternative choices would alter our results. We consider the impact of the timestep size by modelling main sequence-evolved galaxies with progenitors of $M_*=1\times10^{8}\mathrm{M}_\odot$ seeded between $z=0.5$ to $z=6$ in 150~equal bins in cosmic time, equivalent to steps of $50\,\mathrm{Myr}$. As a metric, we determine the deviation in final stellar mass relative to a model run with $\delta t=5\,\mathrm{Myr}$. For timestep sizes of $\delta t=10, 100, 200,$ and $1000 \,\mathrm{Myr}$, the median absolute deviation is $0.0015, 0.027, 0.053$ and $0.207\,$dex ($0.35\%, 6.4\%, 13\%$ and $61\%$), while the maximum difference in stellar mass for any progenitor is $0.003, 0.053, 0.102$ and $0.39\,$dex ($0.7\%, 13\%, 27\%$ and $140\%$). The ratio in final stellar mass is dependent on when the progenitor is seeded, with shorter timebins resulting in increasingly larger final masses for increasingly earlier seeding. We attribute this to the fact that the SFR increases with the stellar mass, so updating the mass to higher values more frequently results in a net increase overall. This difference would likely be more stark if the SFMS itself was not rapidly sinking, such that more frequent updates can, in particular at later times, effectively decrease the SFR and thus final stellar mass. 

Another case to consider is when the SFR varies around the value dictated by the SFMS and the model galaxy stellar mass. We find that adding a lognormal scatter in SFR, $\log{\mathrm{SFR}(M_*,t)} = \log{\mathrm{SFR}_\mathrm{\small{SFMS}}(\mathrm{M_*},t)} + \Delta_\mathrm{SFMS}$ with $\Delta_\mathrm{SFMS}$ drawn from a normal distribution of variance $\sigma(\Delta_\mathrm{SFMS})>0$, necessarily results in larger stellar masses. This can be understood when considering that scattering $0.2$dex above the SFMS SFR results in more net produced stellar mass than subsequently lying $0.2$dex below the SFMS value loses (60$\%$ more vs $40\%$ less). By that same consideration, the net stellar mass gain relative to purely main-sequence-evolved galaxies increases with increasing scatter. 

It is for the above reason that we have preferred here to work with fixed logarithmic separations from the SFMS, $\Delta_\mathrm{SFMS}=const$. We have confirmed that applying a fixed $\Delta_\mathrm{SFMS}$ does not significantly increase the impact of our choice of timesteps. For our maximum $\Delta_\mathrm{SFMS}=+0.5$dex, the median absolute difference in final stellar mass between $\delta t=5\,\mathrm{Myr}$ and our choice $\delta t=100\,\mathrm{Myr}$ is $0.46\,$dex ($11\%$) while the maximum difference is $0.47\,$dex ($11.5\%$). For any negative $\Delta_\mathrm{SFMS}$ the variation in stellar masses for different time bins actually decreases, as model galaxies reach very low sSFRs earlier, which do not significantly alter their stellar mass any more, regardless of timestep.

In summary, we find the median absolute deviation in the stellar masses of our modeled galaxies for a time step size of up to $\delta t=100\,\mathrm{Myr}$ to be around $0.027\,$dex, or $\approx 7\%$, when compared to the stellar masses determined with $\delta t=5\,\mathrm{Myr}$. We therefore conclude that our choice of timestep size $\delta t=100\,\mathrm{Myr}$ does not significantly impact our presented results. In addition, it is worth mentioning that observed star formation rates from various tracers themselves trace different timescales, shorter ($\approx10\,$Myr) for H$\alpha$ and longer ($\approx100\,$Myr) for FUV and NUV bands \citep{kennicutt:2012}. 

\section{Main sequence star formation history fits}
\label{app:sfms_fits}

\begin{figure}
\centering
\includegraphics[width=0.95\columnwidth]{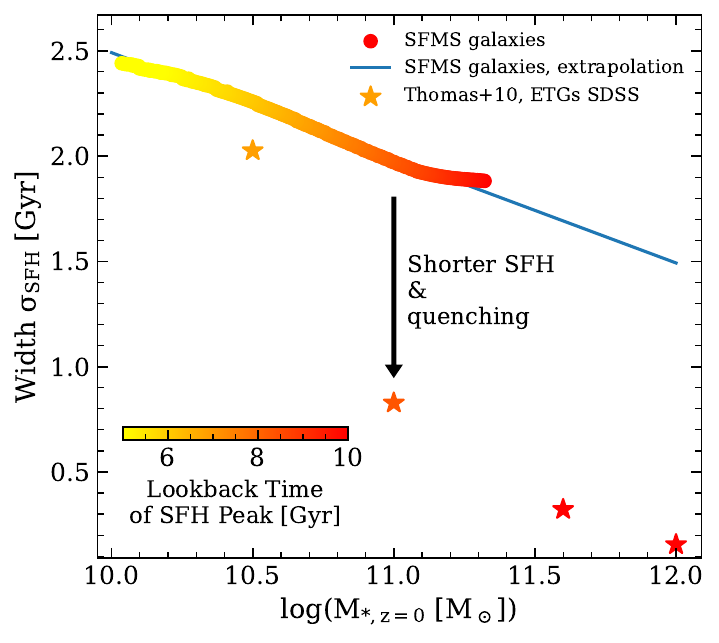}
\caption{The width $\sigma_\mathrm{SFH}$ of a Gaussian fit to the specific star formation rate history of galaxies that evolved fully along the SFMS (shown in the top panel of Fig.~\ref{fig:f_age}), colored by the lookback time of the peak of the Gaussian. Plotted are only models where the progenitor galaxy accounts for less than $1\%$ of the final mass.}\label{fig:fits}
\end{figure}

As discussed for the bottom panel of Fig.~\ref{fig:f_age}, we fit Gaussians to the specific star formation histories of our model galaxies seeded between $z=0.5$ and $z=6$, which were fully evolved along the SFMS. We found a good agreement in the trend of earlier peaks for more massive galaxies when compared to ETGs from SDSS and ATLAS$^\mathrm{3D}$. In Fig.~\ref{fig:fits} we also consider the width of the Gaussian fit, $\sigma_\mathrm{SFH}$, as a function of the final stellar mass. We find for SFMS evolved galaxies that those which end up being more massive exhibit shorter and more peaked SFHs, as also found by \citet{thomas:2010} for SDSS ETGs. However, their SFHs are more peaked, which is to be expected, as they can have quenched. This quenching cuts the SFHs short compared to our SFMS galaxies, and thus reaches a level of peakedness that we by design cannot reproduce. It is still interesting to note that the same mass trend as found for ETGs also appears within SFMS evolved galaxies. 

\section{Connecting present-day galaxies and their progenitors}
\label{app:finding}

\begin{figure*}
\centering
\includegraphics[width=0.9\textwidth]{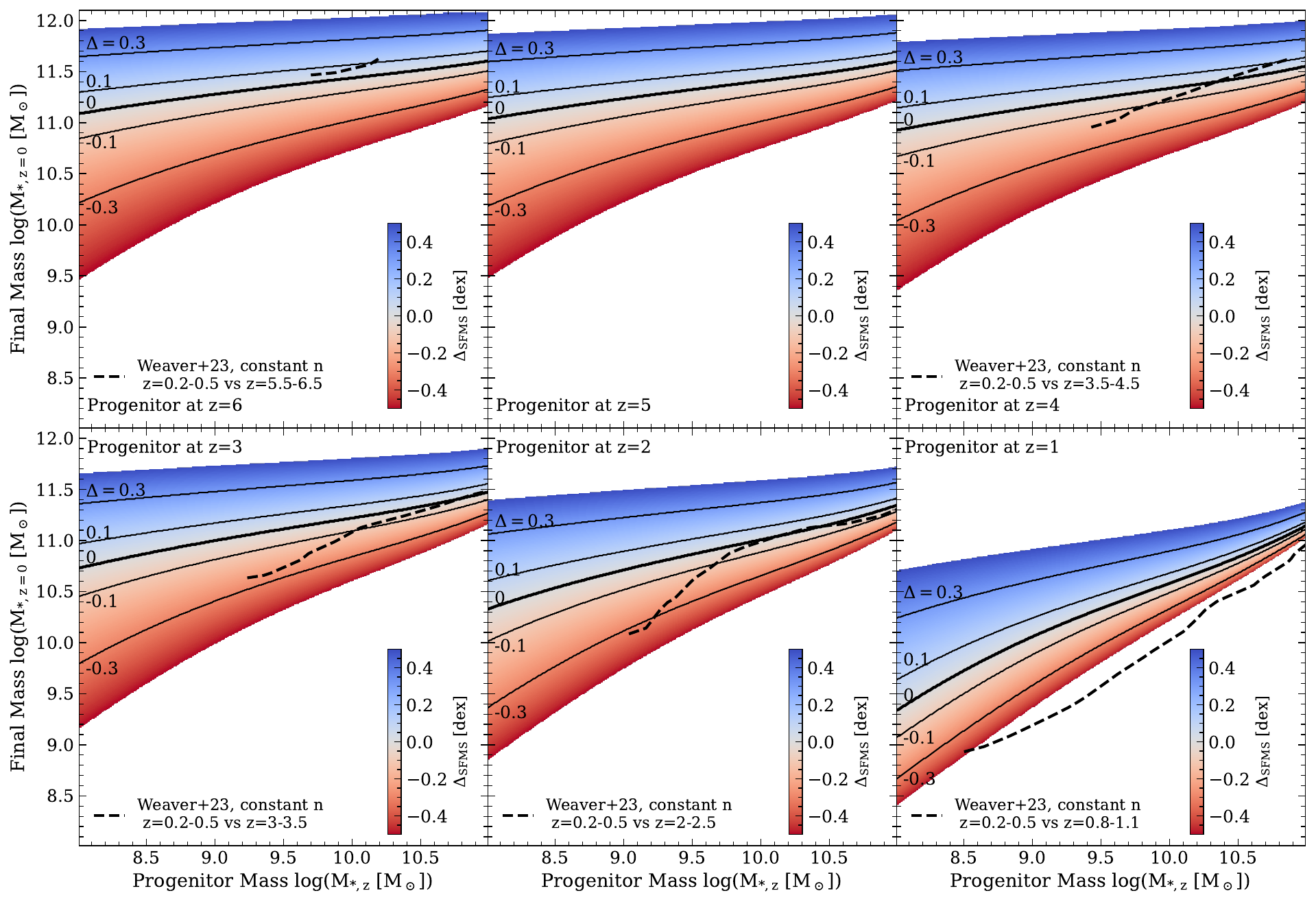}
\caption{As for the right panel of Fig.~\ref{fig:grid_chab}, the present-day galaxy stellar masses $\mathrm{M}_{*,z=0}$ relative to the stellar masses of their progenitor galaxies at redshifts $z=6$ (\textbf{top left}) to $z=1$ (\textbf{bottom right}). The color indicates the logarithmic separation from the SFMS, with blue being galaxies that evolved with SFRs consistently above the SFMS, while red denotes galaxies that evolved consistently below. The dashed lines give the expected connection between present and past galaxy stellar masses under the assumption that their cosmic number density has remained constant, where we use the galaxy stellar mass function determined by \citet{weaver:2023} for the matching. There is no line for $z=5$ as the galaxy stellar mass function is not monotonically decreasing there, leading to different galaxy stellar masses with identical number densities.}\label{figA:grid_chab}
\end{figure*}

We provide in Fig.~\ref{figA:grid_chab} the relation between progenitor stellar mass and final $z=0$ stellar mass for galaxies that evolved on or at some fixed separation in log-space, $\Delta_\mathrm{SFMS}$, from the SFMS. These are as in the right panel of Fig.~\ref{fig:grid_chab}, with the color showing the fixed separation being either above (blue) or below (red) the SFMS SFR while the solid black lines going from top to bottom highlight $\Delta_\mathrm{SFMS}=0.3, 0.1, 0, -0.1,$ and $-0.3\,$dex. The dashed black line is the expected connection between the $z=0$ descendant and $z\geq1$ progenitor galaxies obtained from the galaxy stellar mass function by \citet{weaver:2023} by matching equal comoving number densities across those redshifts. There is no line for $z=5$ as the galaxy stellar mass function is not monotonically decreasing there, resulting in two stellar masses having the same number density. 

We see that for early progenitors ($z=6$, top left) the number density evolution would match expectations from the main sequence. That is to say, if we take galaxies of $M_*=1\times10^{10}\mathrm{M}_\odot$ at $z=6$, the $z=0$ mass of equivalent number density is $M_*=3\times10^{11}\mathrm{M}_\odot$, which is essentially identical to the expected mass if those galaxies had evolved along the SFMS until $z=0$. This remains true at lower redshifts (down to $z=1$) for the case of high mass progenitors $M_*>1\times10^{10}\mathrm{M}_\odot$, meaning that their number densities follow expectations from in-situ SFMS evolution. This is noteworthy as these massive galaxies are known to evolve secularly through mergers, which therefore must bring in a comparable amount of stellar mass to that which would have been produced in-situ if they had remained along the SFMS. For lower mass progenitor galaxies, instead, the $z=0$ descendants of equal comoving number density are less massive than expected from an SFMS evolution. This means that they must have spent some amount of time well below the SFMS $\Delta_\mathrm{SFMS}=-0.3$.

\end{appendix}

\end{document}